# Area Rate Evaluation based on Spatial Clustering of massive MIMO Channel Measurements

Maximilian Arnold, Johannes Pfeiffer and Stephan ten Brink
Institute of Telecommunications, Pfaffenwaldring 47, University of Stuttgart, 70569 Stuttgart, Germany
Email: {arnold,pfeiffer,tenbrink}@inue.uni-stuttgart.de

*Abstract*—Channel models for massive MIMO are typically based on matrices with complex Gaussian entries, extended by the Kronecker and Weichselberger model. One reason for observing a gap between modeled and actual channel behavior is the absence of *spatial consistency* in many such models, that is, spatial correlations over an area in the $x$, $y$-dimensions are not accounted for, making it difficult to study, e.g., area-throughput measures. In this paper, we propose an algorithm that can distinguish between regions of non-line-of-sight (NLoS) and line-of-sight (LoS) via a rank-metric criterion combined with a spiral search. With a $k$-means clustering algorithm a throughput per region (i.e., cluster) can be calculated, leading to what we refer to as "area-throughput". For evaluating the proposed orthogonality clustering scheme we use a simple filtered MIMO channel model which is *spatially consistent*, with known degrees of freedom. Moreover, we employ actual (spatially consistent) area channel measurements based on spatial sampling using a spider antenna and show that the proposed algorithm can be used to estimate the degrees of freedom, and, subsequently, the number of users that maximizes the throughput per square meter.

## I. INTRODUCTION

Channel modeling is one key aspect for fully understanding the opportunities and advantages of massive MIMO. Initially, commonly used MIMO-channel models, like using matrices with independent and identical distributed (i.i.d.) complex Gaussian entries, extended by the Kronecker and Weichselberger model were applied to various massive MIMO scenarios [1], [2]. To verify the viability of those models, channel measurements were conducted with a large number of antennas ([3], [4], [5], [6]). It was shown that a gap exists between the actual channel behavior, and the commonly used MIMO models. To explain this mismatch it was observed through measurements [7] that the Wide Sense Stationary Uncorrelated Scattering (WSSUS) condition over the large array does no longer hold, as a scatterer could be closer to the array than the Rayleigh distance, i.e., the range between the near and far-field of an antenna array.

Owing to these shortcomings, geometry-based channel models were proposed where the WSS condition was modeled via grouping scatterers into clusters [8]. By this, each antenna sees a predefined set of scatterers that can be different from other antennas' view. This concept was added to the Kronecker model in [9], where a birth-death process over the array was employed. Still, even with these more comprehensive channel models, a spatial component is still missing, resembling the "richness" of the scattering environment. This leads to the conclusion that, in any environment with a large number of antennas, a sufficient number of scatterers should be given to close the gap to the Kronecker model, with the only remaining limitation being the antenna correlation.

As shown in [10] it is crucial to compute throughput estimations for such "area" massive MIMO channel models and compare them to predictions using previously known channel models. For this, we employ area channel measurements based on spatial sampling using a spider antenna [11] and propose an algorithm that can distinguish between regions of non-line-of-sight (NLoS) and regions of line-of-sight (LoS) via a rank-metric criterion. In the LoS-case the well-known beamforming patterns show up, with ray-like clusters, while in the NLoS-case a much finer spatial granularity of such orthogonality clusters can be observed. To achieve this, a $k$-means clustering algorithm splits up the entire spatial area under consideration into $k$ different regions. With these regions, a throughput per region (or cluster) is calculated, leading to what we refer to as "area-throughput". By sweeping over the possible number of users $k$ we determine the number of users per square meter that maximizes the area-throughput, dependent on the LoS/NLoS-environment.

For evaluating the proposed clustering algorithm we first introduce a simple filtered MIMO channel model (also referred to as "fake", or "simulated" channel model) which is *spatially consistent*, that is, it also models spatial correlations in the $x$, $y$-dimensions; obviously, for this constructed model, the degree of freedom over the area is known. We show that the proposed algorithm can detect the number of users that maximizes the throughput very well. We also apply the clustering algorithm to spatially sampled area measurements in an office environment as provided by [11] and find similar behavior as for the filtered channel model; however, the limited richness of the measured office environment does not allow users to be as close as $\lambda/2$ for 16 transmit antennas, which the proposed clustering algorithm correctly predicts: it detects the spatial degrees of freedom within a specific region and, with this information, the number of users per square meter can be computed that maximizes the throughput; this further emphasizes the importance of considering spatial consistency in channel modeling.

The paper is organized as follows: Section II describes the channel models used and briefly introduces the spatially sampled area channel measurements. Section III explains the detection of the LoS and NLoS regions, while Section IV proposes a method for calculating the area throughput for any given spatial channel measurement, which, then, is applied to both, the "simulated" filtered channel, and to the actual channel measurements of an indoor office environment. Finally, Section V renders some conclusions.

## II. CHANNEL MODELS AND AREA MEASUREMENTS

In this Section we briefly introduce a method to create *spatially consistent* channel models for the LoS and NLoS case. Also, we introduce spatially-sampled channel measurements obtained using a spider-antenna, which, by their very nature, are spatially consistent.

### A. Simulated: Line-of-Sight (LoS)

For the Line-of-Sight (LoS) case the channel from a basestation to a user can be calculated as

$$h_{\text{LoS}} = \frac{\lambda}{4\pi d} \cdot e^{j2\pi \frac{d}{\lambda}} \quad (1)$$

with $d$ being the distance between user and the respective antenna of the array, and $\lambda$ is the free-space wavelength at the considered carrier frequency $\lambda = c_0/f_c$. With (1) a spatial map of channel coefficients (one per antenna) can be created by calculating the distance between each antenna and the area positions, respectively, e.g., located on a regular grid.

### B. Simulated: Non-Line-of-Sight (NLOS)

To create an NLoS-spatially consistent channel, a spatially sampled map is created using $1 \times M$-vectors with i.i.d. complex Gaussian entries, i.e.,

$$\mathbf{h}_{\text{NLoS,stationary}} \sim \alpha \mathcal{CN}(0, \sigma^2) \quad [1 \times M], \quad (2)$$

where $\alpha$ is the average path-loss from the base-station (BS) to the user, and $M$ is the number of BS antennas, and $\sigma^2 = 1$. Thus, (2) creates a grid of "stationary points" at a distance of $\lambda/2$ in the $x, y$-plane, and each channel (vector) at a specific stationary point is quasi-orthogonal to any other of such stationary point channels. Based on this grid, a spatially consistent channel map is created using a two-dimensional rotational symmetric Gaussian filter $\mathbf{G}(x, y)$ of length $\lambda/2$, which can be mathematically described by a two-dimensional convolution

$$\mathbf{H}_{\text{NLoS,ups.}}(x, y) = \sum_{n=-\infty}^{\infty} \sum_{k=-\infty}^{\infty} \mathbf{H}_{\text{NLoS,zeroPadded}}(n, k) \cdot \mathbf{G}(x - n, y - k)$$

where $\mathbf{H}_{\text{NLoS,zeroPadded}}$ is the zero-padded version of $\mathbf{H}_{\text{NLoS,stationary}}$ with an upsampling factor $L$. The filter is defined as

$$\mathbf{G}(x, y) = \begin{cases} e^{-\frac{|x|^2 + |y|^2}{L^2}} & x \leq L/2, y \leq L/2 \\ 0 & \text{else} \end{cases} \quad (3)$$

Note that the (spatial) sampling theorem is fulfilled, as the length of the filter is smaller than $L$ and, therefore, the stationary points are uncorrelated.

### C. Measured channel with a spider antenna

We briefly introduce the spatial channel measurements conducted in [11]. A spider antenna was used to move a software-defined radio as a transmitter in the $x, y, z$-plane, to measure a spatially consistent channel in three spatial dimensions, and to study linear transmitter precoding.

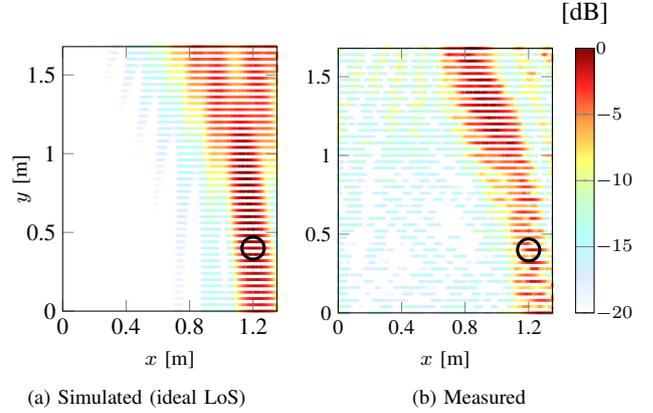

Fig. 1. Spatial energy map, simulation vs spatial area measurement for MR precoding with respect to a target user at $x = 1.2$m, $y = 0.4$m, 16-antenna linear array, LoS scenario.

The Fig. 1 shows the spatial energy, if 16 antennas in a line with a distance of $\lambda/2$ are precoded with respect to the position $x = 1.2$ m and $y = 0.4$ m using a maximum ratio (MR) precoder. Moreover it was verified that the measurements are reproducible and the SNR was always above 20 dB, see [11] for further information. Observe that both precoding maps, based on simulated and measured channel, respectively, match remarkably well. Also, with the same set-up, NLoS channels were measured to better understand the potential of a rich scattering environment in the context of massive MIMO. These measurements, LoS and NLoS, will be further studied later when using $k$-means clustering.

## III. DETECTION OF LOS AND NLOS REGIONS

### A. Clustering based on rank-criterion

Massive MIMO is showing its true potential in an NLoS environment as it can use reflections to separate closely spaced users. On the other hand, massive MIMO boils down to beamforming in the LoS case. This leads to the fact that for performance estimation it is necessary to split the area into spatial regions of LoS and NLoS behavior. For this, we propose a rank-based detection over a spiral search algorithm. The algorithm works as follows: At first, $k$ random positions are taken from the spatially measured area as starting points. Around each starting point a spiral search is kicked off where, with every step, a channel vector is added to a temporary channel matrix $\mathbf{H}_{\text{tmp}}$. The rank of the corresponding channel matrix is calculated, which determines whether the position is added to the area or not. If the rank is increased beyond a certain rank threshold the position is removed from the current area positions. This is done until one full perimeter has been closed around the area without adding new points to the channel.

For robustness against noise the rank of the temporary channel matrix is calculated over the power of the eigenvalues which writes as

$$\text{Rank}(\mathbf{H}_{\text{tmp}}) = \sum_{\forall i} \left\langle \Lambda_i^2 > 0.01 \cdot \sum_j \Lambda_j^2 \right\rangle$$

where each significant eigenvalue $\Lambda_i$ is counted if its power contributes more than $1\,\%$ to the total power, and $i = 1,..,\min(N,M)$. The notation $\langle \cdot \rangle$ denotes the Iverson brackets, which turn each logical operator to 1 if the condition is fulfilled, and to 0 otherwise.

### B. Example of LoS and NLoS separation

To calibrate the performance of this algorithm we create a scenario that contains both, an area with NLoS behavior, and an area with LoS channel behavior. This model writes as

$$\mathbf{h}_{\text{channel}} = \begin{cases} \mathbf{h}_{\text{NLoS}} & x/(20\lambda) > 60,\ y/(20\lambda) > 90 \\ \mathbf{h}_{\text{LoS}} & \text{else} \end{cases}$$

Using the rank-metric based criterion, the spiral search algorithm should detect the regions almost perfectly. As from the theoretical channel model, we expect clusters of beam-like patterns in the LoS region and clusters of circle-like patterns in the NLoS region.

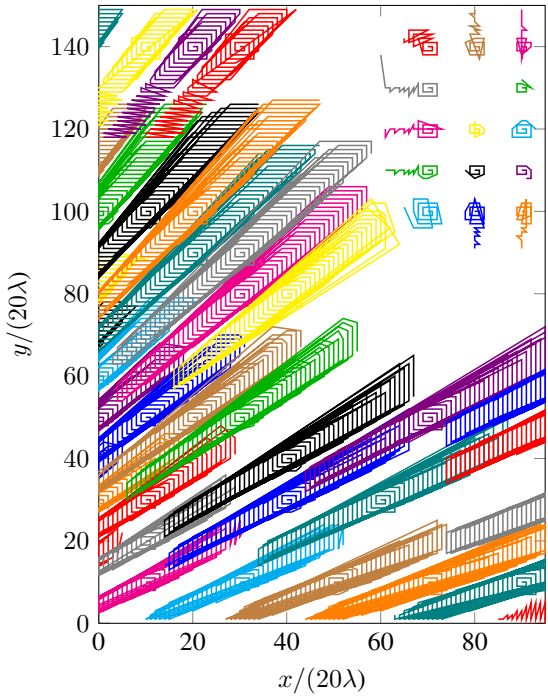

Fig. 2. Simulated: Rank-metric based detection of NLoS and LoS regions

Fig. 2 shows the created regions through the spiral search algorithm, where the rank of the region is one. As expected, the LoS regions show clusters of beam patterns, and the NLoS regions show circle-like patterns. It is instructive to define a circular-to-line ratio $\eta_{\text{NLoS}}$, with

$$\eta_{\text{NLoS}} = 4\pi \frac{\text{Area}}{\text{Perimeter}^2} \qquad \in [0,1]$$

ranging from 0 (i.e., beam-like region, LoS-pattern) to 1 (i.e., circular-like region, indicating "full-scattering" NLoS) that can be used to automatically split spatially consistent measurements into NLoS/LoS-regions. Obviously, from Fig. 2, the algorithm works quite well on the simulated (filtered) channel model. Next we apply the clustering algorithm to an actual channel, i.e., the spider antenna measurements.

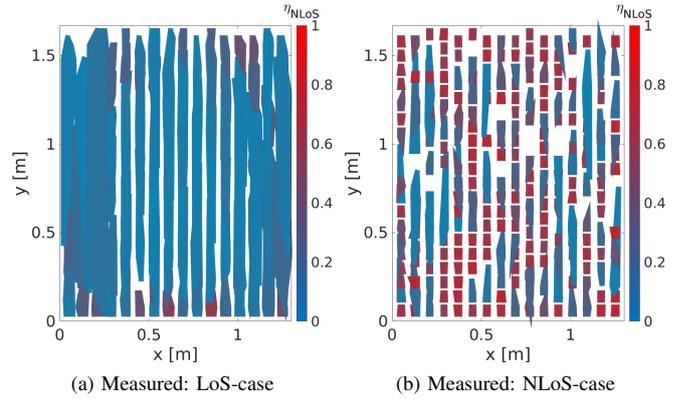

(a) Measured: LoS-case  (b) Measured: NLoS-case

Fig. 3. Rank-metric based detection of LoS and NLoS regions on measured data

Fig. 3 shows the algorithm on the spider antenna measurements in an LoS-case and in an NLoS-case. Again, the beam patterns appear in the LoS-case, and circle/square patterns show up in the NLoS-case, leading to the conclusion that the spiral search algorithm is suitable for distinguishing LoS and NLoS regions on measured channel data, which is a prerequisite for the next steps, as, from now on, we focus on studying NLoS regions.

## IV. QUANTITATIVELY EVALUATING SPATIAL CHANNELS

For evaluating spatial channels and their performance we propose a three-step algorithm for calculating the maximum number of users and, correspondingly, the area throughput for a given scenario.

### A. Step 1: k-means clustering of the area

For a given channel matrix $\mathbf{H} \in \mathbb{C}^{N \times M}$ and corresponding $N$ spatial user positions, the $k$-means clustering calculates the channel centroids for $k$ given users. Initially, a random set of $k$ channels is used as channel centroids

$$\mathbf{h}_{\text{mean},i} = \mathbf{H}_{i,m} \qquad \forall m = 1,..,M$$

where $i$ is an index to a random set of length $k$, see (4). This initializes the $k$-means algorithm. With the first centroids (step $t=1$) we calculate the distance of each channel $\mathbf{h}_n$ to the centroids $\mathbf{h}_{\text{mean},j}$ and find the minimal distance

$$\mathbb{S}_i^t = \left\{ \mathbf{h}_n : i = \arg\min_j \|\mathbf{h}_n - \mathbf{h}_{\text{mean},j}\|^2 \right\} \quad j = 1,..,k \quad (4)$$

where $n = 1,..,N$. After re-sorting, the new centroid per each cluster $i$ is calculated using

$$\mathbf{h}_{\text{mean},i} = \frac{1}{|\mathbb{S}_i^t|} \sum_{\mathbf{h}_n \in \mathbb{S}_i^t} \mathbf{h}_n. \quad (5)$$

With the new centroids $\mathbf{h}_{\text{mean},i}$ the new assignment through (4) can be calculated and, subsequently, the algorithm calculates the next step $t = 2$. This loop continues until the result is assumed to have converged, with condition

$$\mathbb{S}_i^{t+1} = \mathbb{S}_i^t \quad \forall i$$

which means that the assignment does not change anymore.

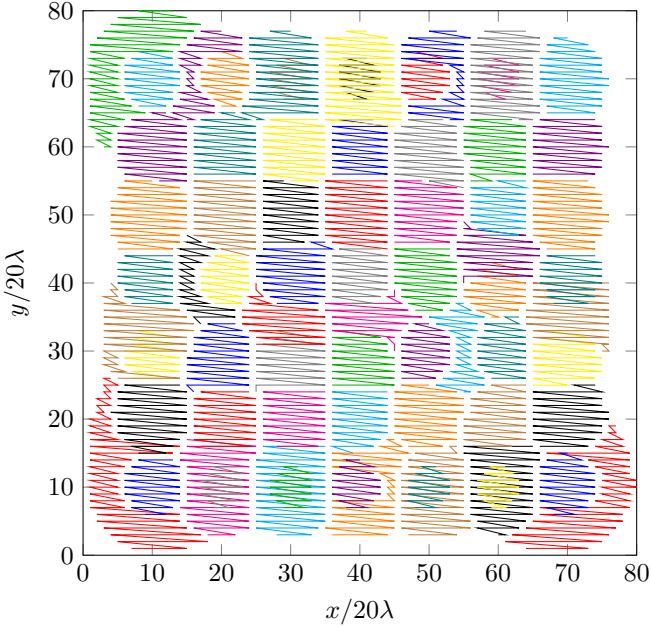

Fig. 4. Simulated: k-means clustering, filtered channel model, $k = 70$

Fig. 4 shows the result of the $k$-means clustering of a simulated/filtered channel created with 7×7 stationary points and $\alpha = 1$, compare to (2), and filtered with a 2D-Gaussian impulse, see (3). An oversampling factor of $L = 20$ is used. Obviously, the result of the clustering is quite close to an expected "checkerboard"-like pattern, around the 49 stationary points on the regular grid. Based on the random initialization of the $k$-means search, some more clusters were created, as the algorithm tries to select the regions "as orthogonally as possible"; yet, keeping in mind that the algorithm has no a priori knowledge on the actual positions of the stationary points, the pattern looks quite plausible. With this result, an average sum-rate per area can be calculated, as will be studied next, i.e., in Step 2.

## B. Step 2: MR-precoding on cluster centroids and sum-rate calculation

The calculated centroids (5) are now used for an MR-precoding matrix [12]

$$\mathbf{P}_{\text{MR}} = \left[ \mathbf{h}_{\text{mean},1}^H, \quad ... \quad , \mathbf{h}_{\text{mean},k}^H \right],$$

where $\mathbf{h}^H$ is the hermitian (i.e., complex conjugate transposed) of the vector $\mathbf{h}$. The received power over the area can be calculated as

$$\mathbf{y}_{\text{RX}} = \mathbf{H}\mathbf{P}_{\text{MR}}\mathbf{x},$$

where $\mathbf{x}$ is the $M \times 1$ transmit vector and $\mathbf{y}_{\text{RX}}$ is the $N \times 1$ receive vector. With this definition, the SIR per cluster can be computed as

$$\mathbf{SIR}_i = \frac{\left\|\mathbf{h}_n \mathbf{h}_{\text{mean},i}^H\right\|^2}{\sum_{j=1,\,j\neq i}^{k} \left\|\mathbf{h}_n \mathbf{h}_{\text{mean},j}^H\right\|^2} \quad \forall \mathbf{h}_n \in \mathbb{S}_i,$$

where $\mathbf{SIR}_i$ is a vector of dimension $1 \times |\mathbb{S}_i|$. To determine the throughput within a spatial area, a sum-rate for the given cluster points needs to be calculated using

$$\mathbf{R}_i(\ell) = \log_2\left(1 + \mathbf{SIR}_i(\ell)\right) \quad \ell = 1,..,|\mathbb{S}_i|.$$

Note that there are points in a cluster with both high and low SIR, as can be seen from Fig. 5 for a cluster at position $x/(20\lambda) = 106$ and $y/(20\lambda) = 93$.

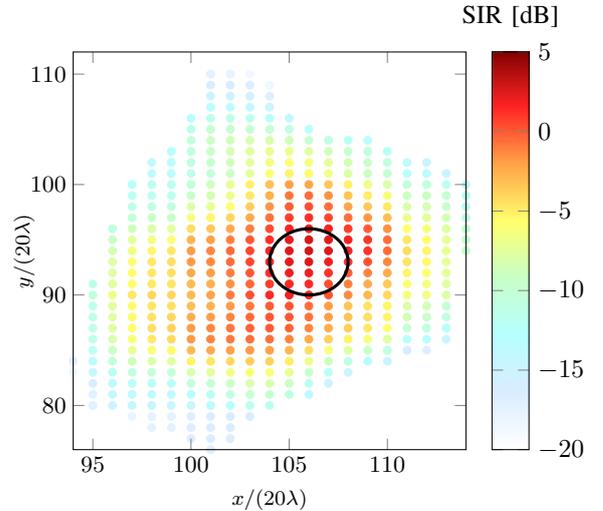

Fig. 5. Example, spatial distribution of the SIR within a cluster at $x/(20\lambda) = 106$, $y/(20\lambda) = 93$

With this, a $q$-percentile can be defined, denoting the $q$ percent users within the cluster having the highest SIRs (e.g., the points inside the circle in Fig. 5 mark the 5%-percentile). Thus, the sum-rate for a given $q$-percentile computes as

$$R_{i,q} = \mathbf{R}_{i,\text{sorted}}\left(\lfloor q|\mathbb{S}_i|\rfloor\right),$$

where $\mathbf{R}_{i,\text{sorted}}$ is the sorted vector of $\mathbf{R}_i$ in ascending order. To determine the area throughput, the effective sum-rate in bits per channel use (bpcu) at a given percentile is determined using

$$R_{q,\text{perCluster}} = \frac{\sum_i^k R_{i,q}}{k} \quad \left[\frac{\text{bpcu}}{\text{cluster}}\right].$$

With this, an area throughput (as shown in [13]) can be calculated as

$$R_q = D\left[\frac{\text{cluster}}{\text{m}^2}\right] \cdot R_{q,\text{perCluster}}\left[\frac{\text{bpcu}}{\text{cluster}}\right]$$
$$= D \cdot R_{q,\text{perCluster}} \quad \left[\frac{\text{bpcu}}{\text{m}^2}\right],$$

where $D$ is the cluster density. Next, we apply this algorithm to the simulated/filtered channel model and the measurements.

### C. Step 3: Determine the behavior of a spatial channel

*1) Simulated Channel:* Note that, for this evaluation, the correlation of spatial channels only stems from the environment geometries and not, like given in the Kronecker-model, from TX and RX antenna correlations (comp. [14]). Thus, we consider the richness of an environment, and not the quality of a specific antenna geometry.

In a first step, the example from the Fig. 4 is used to show the capabilities of this model. As a second step, the algorithm is applied to the spider antenna measurements in the NLoS case.

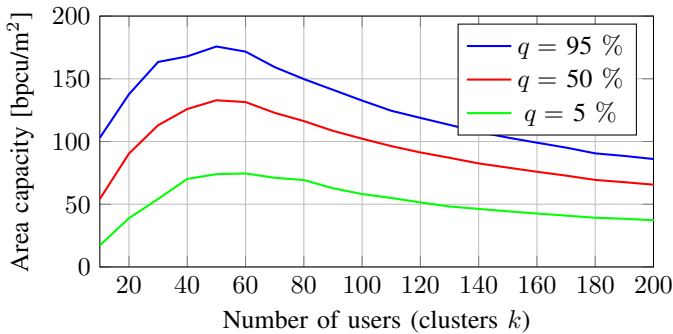

Fig. 6. Simulated: Area sum-rate for different SIR-percentiles with a filtered channel model for 64 antennas

Fig. 6 shows the area sum-rate versus the number of possible users for different SIR-percentiles. Note that the number of users is given by the corresponding $k$ clusters. As can be seen, the different SIR-percentiles behave pretty much the same, and thus, it is sufficient to just look at the median (50%) percentile. By increasing the number of users (clusters), the area throughput first increases, due to the better orthogonalization of different user in this area. After the number of stationary points in this environment based on the number of TX antennas is reached, the interference due to non-orthogonal clusters increases and limits the system performance. Moreover, as a maximum for the given area, the stationary points of 49 clusters shows up. This means that the degrees of freedom, which were used to create this channel model in the first place, can be reliably identified, as the maximal area throughput is reached at the same number of clusters. Still, an open question remains pertaining the influence of the number of antennas on the maximum number of users in an area. For this, we perform a parameter sweep over the number of transmit antennas.

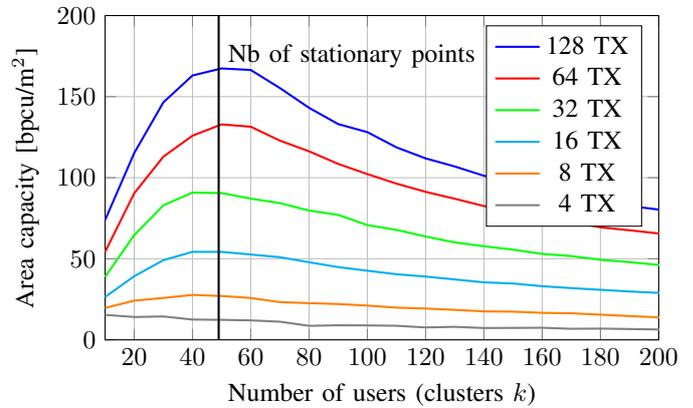

Fig. 7. Simulated: Area sum-rate for different number of transmit antennas with a filtered channel model and median SIR-percentile

The dependency of the area throughput on the number of transmit antennas is shown in Fig. 7. It can be seen, also, that the area throughput for all number of antennas first increases, until the maximum possible orthogonalization is achieved; then again, the interference starts to limit the system performance. Moreover, it is shown that with increasing number of transmit antennas the maximum area throughput increases and the number of possible users increases until the area is "saturated". This is a key observation for a high number of transmit antennas.

*2) Measured Channel:* Now the spider antenna measurements of [11] for the the NLoS-case are considered.

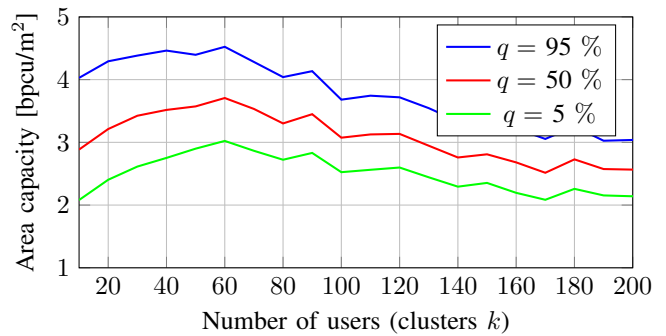

Fig. 8. Measured: Area sum-rate for different SIR-percentiles for a measured channel with 8 TX antennas

The Fig. 8 shows the same behavior as the simulated channel. At first, with increasing number of clusters, the area throughput increases due to orthogonal clusters and, after a saturation point, the area throughput is limited. Moreover the different SIR-percentiles exhibit the same behavior.

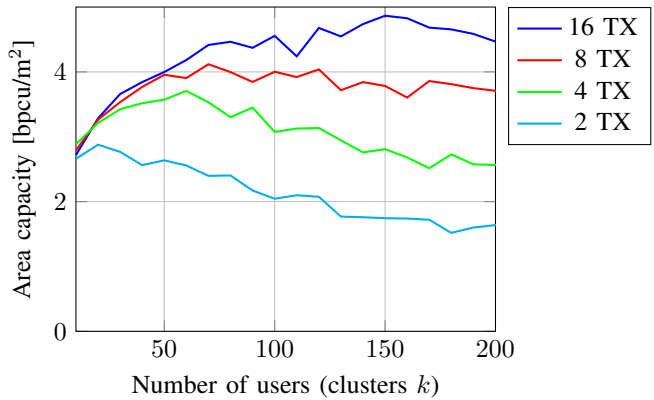

Fig. 9. Measured: Area sum-rate for different number of transmit antennas with a measured channel and median (i.e.,50 %) SIR-percentile

Fig. 9 shows area throughputs for different number of transmit antennas. As can be seen (compare to Fig. 7) the average area throughput increases with more transmit antennas, and the maximum number of users per square meter is increased. This corresponds well with the simulations using a filtered channel model as shown in Fig. 7. Moreover, a maximum is reached for different number of antennas since the clusters are not mutually orthogonal to each other. Observe that the measurement was obtained from an office environment (no rich scattering), and thus, the maximum number of users is not the number of $\lambda/2$-circles which tessellate the area. This means that for a given number of transmit antennas the correlation of $\lambda/2$-spaced users is not sufficient to model a channel, since the richness of the environment can also limit the system performance and, therefore, the area throughput. This leads to the conclusion that, for a finite number of antennas, a spatial component should be added to the channel models, modeling the "richness" of scatterers within the investigated environment in a spatially consistent way.

## V. Conclusion

In this paper, we proposed an algorithm that can classify simulated or measured MIMO data into non-line-of-sight (NLoS) and line-of-sight (LoS) regions, respectively, using a rank-metric based criterion combined with a spiral search. A $k$-means orthogonality clustering algorithm allowed to compute an area-throughput measure that we evaluated using a simple filtered, spatially consistent MIMO channel model. In addition, the clustering algorithm was applied to actual area channel data that was obtained from measurements using a spider antenna. It was shown that the proposed algorithm accurately estimates the degrees of freedom as well as the number of users for maximizing the throughput per square meter.